\documentclass[prb,showpacs, twocolumn, aps,superscriptaddress,floatfix]{revtex4-1}
\usepackage{amsmath}
\usepackage{amssymb}
\usepackage{bm}
\usepackage{graphicx}
\usepackage{color}
\usepackage[svgnames]{xcolor}
\usepackage{soul}
\usepackage{hyperref}
\usepackage{verbatim}

\begin{document}

\title{Half-metal and other fractional metal phases in doped AB bilayer
graphene}

\author{A.L. Rakhmanov}
\affiliation{Institute for Theoretical and Applied Electrodynamics, Russian
Academy of Sciences, 125412 Moscow, Russia}

\author{A.V. Rozhkov}
\affiliation{Institute for Theoretical and Applied Electrodynamics, Russian
Academy of Sciences, 125412 Moscow, Russia}

\author{A.O. Sboychakov}
\affiliation{Institute for Theoretical and Applied Electrodynamics, Russian
Academy of Sciences, 125412 Moscow, Russia}

\author{Franco Nori}
\affiliation{Center for Quantum Computing and Cluster for Pioneering
Research, RIKEN, Wako-shi, Saitama, 351-0198, Japan}
\affiliation{Department of Physics, University of Michigan, Ann
Arbor, MI 48109-1040, USA}

\begin{abstract}
We theoretically argue that, in doped AB bilayer graphene, the
electron-electron coupling can give rise to the spontaneous formation of
fractional metal phases. These states, being generalizations of a more
common half-metal, have a Fermi surface that is perfectly polarized not
only in terms of a spin-related quantum number, but also in terms of the
valley index. The proposed mechanism assumes that the ground state of
undoped bilayer graphene is a spin density wave insulator, with a finite
gap in the single-electron spectrum. Upon doping, the insulator is
destroyed, and replaced by a fractional metal phase. As doping increases,
transitions between various types of fractional metal (half-metal,
quarter-metal, etc.) are triggered. Our findings are consistent with recent
experiments on doped AB bilayer graphene, in which a cascade of phase
transitions between different isospin states was observed.
\end{abstract}


\date{\today}
\maketitle

\section{Introduction}

A usual metal demonstrates perfect symmetry with
regard to the carriers' spin projection. This symmetry manifests itself in
the vanishing total spin magnetization and the Fermi-surface spin
degeneracy. Yet the symmetry can be spontaneously destroyed by sufficiently
strong electron-electron interaction, which may result, for example, in the
formation of two non-identical Fermi surfaces for the two spin projections.
In the extreme case of the so-called half-metals (HM), one of these
projections is completely absent from the Fermi surface, while all states
at the Fermi energy have identical spin quantum
number~\cite{first_half_met1983,half_met_review2008,hu2012half}.
Various rather dissimilar materials with transition-metal atoms are found
to be
half-metals~\cite{nimnsb_exp1990,lasrmno_half_met_exp1998,
cro2_half_met_exp2001,co2mnsi_half_met_exp2014}.
Several
papers~\cite{metal_free_hm2012,meta_free_hm2014,son2006half,kan2012half,
huang2010intrinsic}
predicted the half-metallicity in carbon-based systems as well. The
existence of spin-polarized currents in such systems makes them promising
materials for applications in
spintronics~\cite{review_spintronics2004,hu2012half}.

Graphene-based bilayer and multi-layer systems posses additional quantum
number, the valley index. In these materials, besides the spin-related
polarization, a many-body state may demonstrate a valley polarization.
Therefore, for graphene-based materials, the notion of a HM can be
generalized to include the possibility of a Fermi surface with perfect
valley polarization as well. Such a proposal was put forward in
Ref.~\onlinecite{aa_quarter_met2021},
where the concept of a quarter-metal (QM) was formulated. A Fermi surface
of a QM state is perfectly polarized both in valley and in spin-related
indices. Furthermore, the latter paper explained that both an HM and a QM
should be viewed as specific instances of a more general notion, `a
fractional metal' (FraM). This many-body phase may be realized in materials
with degenerate Fermi surface.
The higher the degeneracy, the stronger fractionalization of the Fermi
surface can be achieved.

Since our
publication~\cite{aa_quarter_met2021}
the experimental observation of a QM state in graphene trilayer has been
claimed~\cite{trilayer_quarter2021exper}.
The experimental data of
Ref.~\onlinecite{ab_frac2022exper}
suggest that a QM and FraM states can be stabilized in a sample of AB
bilayer graphene (AB-BLG). Given these experimental successes it appears
important to develop a microscopic theoretical framework that can explain
the existence of the FraM in the AB-BLG. In this paper, a suitable
mechanism is proposed and discussed.

\section{Model}

An elementary unit cell of the AB-BLG consists of four
atoms (sublattices
$A$
and
$B$,
and layers 1 and 2) with the distance
between neighboring carbon atoms
$a_0\approx0.142$\,nm and interlayer
distance
$c_0\approx 0.335$\,nm. The hoping amplitude
$t$
connecting the
nearest
$A$
and
$B$
sites in the layer is
2.5\,eV~$\lesssim t \lesssim 3$\,eV.
The hopping between the nearest sites in different layers can be estimated
as
0.3\,eV~$\lesssim t_0 \lesssim 0.4$\,eV.
It is possible to introduce additional, longer-range, hopping amplitudes
into the model. We assume, however, that the effect of these amplitudes is
weak, and they are neglected.

The AB-BLG Brillouin zone is a regular hexagon, with two non-equivalent
Dirac points at
\begin{eqnarray}
\mathbf{K}_1 = \frac{2\pi}{3\sqrt{3} a_0} (\sqrt{3}, 1)
\quad
\text{and}
\quad
\mathbf{K}_2 = \frac{2\pi}{3\sqrt{3} a_0} (\sqrt{3}, -1).
\end{eqnarray}
It is convenient to measure momentum relative to the Dirac points. Thus, we
introduce
$\mathbf{q}=\mathbf{k}-\mathbf{K}_{1,2}$.

The energy spectrum of undoped AB-BLG consists of four bands, two electron
and two hole ones. Since we are interested in the low-energy spectrum of
AB-BLG,
$q\ll 2t_0/3ta_0$,
we restrict our consideration to the effective two-band model. It has one
electron and one hole band, both bands have quadratic dispersion. The bands
touch at the Fermi energy. When the (small) trigonal warping terms are
ignored, the Hamiltonian for a single-electron wave function
reads~\cite{mccann_falko2006,jung_mcdonald2014tb,bilayer_review2016}
\begin{eqnarray}
\label{Ham_1}
H_0 =
-\frac{\hbar^2 v_{\rm F}^2}{t_0}
\left(
    \begin{array}{cc}
      0 & (iq_x+\xi q_y)^2 \\
      (iq_x-\xi q_y)^2 & 0 \\
    \end{array}
  \right),
\end{eqnarray}
where the graphene Fermi velocity is
$v_\textrm{F}=3a_0t/2\hbar$
and $\xi$ is the valley index. The value
$\xi=1$
corresponds to
$\mathbf{K}_1$
and
$\xi=-1$
corresponds to
$\mathbf{K}_2$.
In the second-quantization formalism we can write
\begin{eqnarray}
H_0 = \sum_{{\bf q} \sigma \xi l}
	\varepsilon^{\vphantom{\dagger}}_{{\bf q} l}
	\gamma^\dag_{\mathbf{q}l\sigma\xi}
	\gamma^{\vphantom{\dagger}}_{\mathbf{q}l\sigma\xi},
\end{eqnarray}
where the spin projection is denoted by $\sigma$, the index $l$
labels the electron
($l=1$)
or hole
($l=2$)
band, and
$\gamma_{\mathbf{q}l\sigma\xi}$
is the corresponding second quantization operator. The eigenenergies
$\varepsilon_{{\bf q} l}$
of the Hamiltonian~\eqref{Ham_1} are
\begin{equation}
\label{Energy_0}
\varepsilon_{{\bf q} l}= (-1)^{l+1}\frac{\hbar^2v_\textrm{F}^2}{t_0}q^2.
\end{equation}
%
Next we include the electron-electron repulsion into the model. The latter
is a highly non-trivial task. Clearly, the low-energy two-band effective
model~(\ref{Ham_1})
is incompatible with the bare Coulomb repulsion. Instead, an effective
interaction Hamiltonian must be derived. Unfortunately, a compact
description of such an effective interaction remains an elusive theoretical
goal. Indeed, due to multiple factors affecting the many-body physics in
graphene and graphene-based systems, an effective interaction term is quite
complex, with multiple coupling constants, whose non-universal values are
poorly known~\cite{Hwang2008, lemonic_rg_nemat_long2012,
vafek_rg2010,vafek_nemat_rg2010, cvetkovic_multi2012}.
In this situation we prefer to adopt a semi-phenomenological approach,
keeping only the terms that directly contribute to the spin-density wave
(SDW) ordering. It is possible to identify three types of such terms. The
first term arises due to the forward-scattering
\begin{eqnarray}
\label{H_int_0}
\!\!H_{\rm int}^{\rm f}\!
=
\!\frac{V_C}{N_c}\!\!
\sum_{\mathbf{k}\mathbf{k}',ll'\atop \sigma\sigma',\xi\xi'}
\!\!\!\!
	\gamma^\dag_{\mathbf{k}l\sigma\xi}
	\gamma^{\vphantom{\dagger}}_{\mathbf{k}'l\sigma\xi}
	\gamma^\dag_{\mathbf{k}'l'\sigma'\xi'}
	\gamma^{\vphantom{\dagger}}_{\mathbf{k}l'\sigma'\xi'},
\end{eqnarray}
where
$N_c$
is the number of unit cells in the sample, and
$V_C$
is an effective interaction constant whose value can be potentially
extracted from the low-temperature data~\cite{Feldman2009, Martin2010,
Weitz2010, Mayorov2011, Bao2012,Freitag2012, Freitag20122053, Velasco2012,
veligura2012, freitag2013}
on spontaneous symmetry breaking in AB-BLG. The forward scattering is
characterized by a small momentum transfer
$|{\bf k} - {\bf k}'| \ll |{\bf K}_1 - {\bf K}_2|$,
and preserves the band indices
$l$
and
$l'$
of the two participating electrons. Next, one can define the backscattering term
\begin{eqnarray}
\label{eq::BSK}
\!\!H_{\textrm{int}}^{\rm b}\!
=
\!\frac{V_C^{\rm b}}{N_c}\!\!
\sum_{\mathbf{k}\mathbf{k}',ll'\atop \sigma\sigma',\xi}
\!\!\!\!
	\gamma^\dag_{\mathbf{k}l\sigma\xi}
	\gamma^{\vphantom{\dag}}_{\mathbf{k}'l\sigma\bar{\xi}}
	\gamma^\dag_{\mathbf{k}'l'\sigma'\bar{\xi}}
	\gamma^{\vphantom{\dag}}_{\mathbf{k}l'\sigma'\xi},
\end{eqnarray}
where a bar on top of a binary-valued index implies the inversion of the index value (for example, if
$\xi=1$
then
$\bar{\xi} = -1$). For
$H_{\textrm{int}}^{\rm b}$
the transferred momentum is large
$|{\bf k} - {\bf k}'| \sim |{\bf K}_1 - {\bf K}_2|$,
thus we can assume that
$V_C^{\rm b} \ll V_C$.
Finally, the umklapp-type interaction
\begin{eqnarray}
\label{eq::UMK}
\!\!H_{\textrm{int}}^{\rm u}\!
=
\!\frac{V_C^{\rm u}}{N_c}\!\!
\sum_{\mathbf{k}\mathbf{k}',\atop \sigma\sigma',\xi\xi'}
\!\!\!\!
	\gamma^\dag_{\mathbf{k}1\sigma\xi}
	\gamma^{\vphantom{\dagger}}_{\mathbf{k}'2\sigma\xi}
	\gamma^\dag_{\mathbf{k}'1\sigma'\xi'}
	\gamma^{\vphantom{\dagger}}_{\mathbf{k}2\sigma'\xi'}
+ {\rm h.c.},
\end{eqnarray}
represents scattering events in which both electrons change their bands. It
accounts for the coupling between inter-layer dipole moments, which is also
weaker than the coupling between charge densities represented by
$H_{\textrm{int}}^{\rm f}$.
In principle, there is backscattering umklapp, which we do not consider due
to it being even weaker than
$H_{\textrm{int}}^{\rm u}$.

\section{Mean-field approximation}

We consider a zero-temperature SDW
instability of the AB-BLG. This is characterized by the spontaneous
generation of staggered spin magnetization violating the spin-rotation
symmetry. The direction of this magnetization is not fixed and there are
several equivalent choices for an SDW order parameter that differ by the
spin-magnetization direction. It is convenient to assume that
$\langle
	\gamma^\dag_{\mathbf{k}1\sigma\xi}
	\gamma^{\vphantom{\dagger}}_{\mathbf{k}2\bar{\sigma}\xi}
\rangle
\ne 0$.
This choice corresponds to the magnetization in the $xy$-plane. Note also
that the introduced order parameter accounts for the coupling of
single-electron states in the same valley $\xi$.

Now, assuming that the
backscattering~(\ref{eq::BSK})
and the
umklapp~(\ref{eq::UMK})
are weak, we apply the mean-field approximation to
$H^{\rm f}_{\rm int}$
\begin{equation}
\label{H_MF}
H_{\textrm{int}}^{\textrm{MF}}
=
-\sum_{\mathbf{k}\sigma\xi}
	\Delta_{\sigma\xi}
	\gamma^\dag_{\mathbf{k}2\sigma\xi}
	\gamma^{\vphantom{\dagger}}_{\mathbf{k}1\bar{\sigma}\xi}
+
{\rm h.c.} + B,
\end{equation}
where the order parameter
$\Delta_{\sigma \xi}$
and
$c$-number
$B$
are
\begin{eqnarray}
\label{Delta_definition}
\Delta_{\sigma\xi}&=&\frac{V_C}{N_c}
\sum_{\mathbf{q}}
\langle
	\gamma^\dag_{\mathbf{q}1\sigma\xi}
	\gamma^{\vphantom{\dagger}}_{\mathbf{q}2\bar{\sigma}\xi}
\rangle\Theta(q_C-q),\\
B&=&
\sum_{\mathbf{q}\sigma \xi}
	\Delta_{\sigma\xi}
	\langle
		\gamma^\dag_{\mathbf{q}2\sigma\xi}
		\gamma^{\vphantom{\dagger}}_{\mathbf{q}1\bar{\sigma}\xi}
	\rangle \Theta (q_C - q)
= \frac{N_c}{V_C} \sum_{\sigma \xi} |\Delta_{\sigma \xi}|^2.\nonumber\\
\label{B}
\end{eqnarray}
In these expressions, the momentum cutoff for the interaction
$q_C$
satisfies
$q_C\ll|\mathbf{K}_1-\mathbf{K}_2|$.

The mean-field
Hamiltonian~(\ref{H_MF})
does not conserve spin (spin-rotation symmetry is spontaneously broken for
non-zero
$\Delta_{\sigma \xi}$).
However, quasi-momentum
${\bf q}$
is conserved. In addition to
${\bf q}$,
one can introduce valley and spin-flavor operators
\begin{equation}
\label{eq::Sfv_def}
S^{\rm f}_{\bf q}
=\!\!
\sum_{\sigma \xi l }
		(-1)^{l+1}\sigma
		\gamma^\dag_{\mathbf{q}l\sigma\xi}
		\gamma^{\vphantom{\dagger}}_{\mathbf{q}l\sigma\xi},
\ \
S^{\rm v}_{\bf q}
=\!\!
\sum_{\sigma \xi l }
		\xi
		\gamma^\dag_{\mathbf{q}l\sigma\xi}
		\gamma^{\vphantom{\dagger}}_{\mathbf{q}l\sigma\xi},
\end{equation}
which commute with the Hamiltonian
$H_0 + H^{\rm MF}_{\rm int}$
and are good quantum numbers. Thus, in this approximation all fermionic
degrees of freedom can be grouped into four uncoupled sectors, each sector
having its own values of spin-flavor index
$(-1)^{l+1} \sigma$
and valley index $\xi$. A sector is characterized by its own order
parameter
$\Delta_{\sigma \xi}$,
and single-particle spectrum
\begin{equation}
\label{MF_energy}
E^{1,2}_{\mathbf{q}\sigma\xi}
=
\pm\sqrt{\Delta^2_{\sigma\xi}+\left(\frac{\hbar^2v_\textrm{F}^2}{t_0}\right)^2q^4}.
\end{equation}
The thermodynamic grand potential
$\Omega$
can be expressed as a sum
\begin{eqnarray}
\Omega = \sum_{\sigma \xi} \Omega_{\sigma \xi} + B,
\end{eqnarray}
where
$\Omega_{\sigma \xi}$
are four partial grand potentials corresponding to specific sectors. At
zero temperature, these are
\begin{equation}
\Omega_{\sigma \xi}
=
\sum_{\mathbf{q}l}
	\left(E^l_{\mathbf{q}\sigma\xi}-\mu\right)
	\Theta\left(\mu-E^l_{\mathbf{q}\sigma\xi}\right),
\end{equation}
where $\mu$ is the chemical potential.

Minimization of $\Omega$ over the order parameters allows us to derive the
following independent self-consistency equations for the order parameters
in the four sectors
\begin{eqnarray}
\label{minimization}
1 = \frac{V_C}{N_c}
	\sum_{|\mathbf{q}|<q_C}
		\frac{\Theta(\mu+E^1_{\mathbf{q}\sigma\xi})
			-\Theta(\mu-E^1_{\mathbf{q}\sigma\xi})}
		{E^1_{\mathbf{q}\sigma\xi}}.
\end{eqnarray}
Since the model is electron-hole symmetric, we can limit our discussion to
the
$\mu>0$
case only. For positive chemical potential:
$\Theta(\mu+E^1_{\mathbf{q}\sigma})-\Theta(\mu-E^1_{\mathbf{q}\sigma})
=\Theta(E^1_{\mathbf{q}\sigma}-\mu)$.
Introducing dimensionless variables
\begin{equation}
\label{dim_var}
g=\frac{V_Ct_0}{\sqrt{3}\pi t^2},\,\,
m=\frac{4t_0\mu}{9t^2},\,\,
\delta_{\sigma\xi}=\frac{4t_0\Delta_{\sigma\xi}}{9t^2},
\end{equation}
we obtain from Eq.~\eqref{minimization}
\begin{equation}
\label{Delta}
1=2g\int^{Q_C}_{Q^m_{\sigma\xi}} \frac{QdQ}{\sqrt{\delta^2_{\sigma\xi}+Q^4}},
\end{equation}
where
\begin{eqnarray}
Q_C=a_0q_C,
\quad
Q^m_{\sigma\xi}=(m^2-\delta_{\sigma\xi}^2)^{1/4}.
\end{eqnarray}
It is evident that the gap in the spectrum of electrons in the sector
$(\sigma,\xi)$
arises only if
$Q_C>Q^m_{\sigma\xi}$,
that is, if the number of the doped charge carriers in this sector is not too large. One can perform the integration in Eq.~\eqref{Delta} and obtain that
\begin{equation}
1=g\ln \left(
	\frac{Q_C^2+\sqrt{\delta^2_{\sigma\xi}+Q_C^4}}
		{m+\sqrt{m^2-\delta_{\sigma\xi}^2}}
\right).
\end{equation}
In the weak coupling limit,
$g\ll 1$,
we have
$\delta_{\sigma \xi} \ll Q_C^2$.
Consequently
\begin{equation}
\label{Delta_xi}
\Delta_{\sigma\xi}=\sqrt{\Delta_0(2\mu-\Delta_0)},
\end{equation}
where
\begin{equation}
\label{Delta_0}
  \Delta_0=\frac{9t^2}{4t_0}q_C^2a_0^2e^{-1/g}
\end{equation}
is the mean-field gap of undoped AB-BLG. Further defining
\begin{eqnarray}
\delta_0 = \frac{4t_0\Delta_0}{9t^2},
\end{eqnarray}
we can express Eq.~(\ref{Delta_xi}) in dimensionless form
\begin{eqnarray}
\delta_{\sigma\xi}=\sqrt{\delta_0(2m-\delta_0)}.
\end{eqnarray}
For finite doping,
Eq.~(\ref{Delta_xi})
implies that the chemical potential must satisfy
$\mu \geq \Delta_{\sigma \xi}$.
Such a relation is naturally expected: to start doping, the chemical
potential must exceed the gap.

Since experiments are performed at fixed doping, we need to connect the values of
$\Delta_{\sigma \xi}$
with doping. It is convenient to introduce partial doping, that is, the
number of electrons with specific values of
$(-1)^{l+1} \sigma$
and $\xi$:
\begin{eqnarray}
\label{doping_full}
x_{\sigma \xi}=-\frac{\partial\Omega_{\sigma \xi}}{\partial\mu}
=
\frac{2\pi}{V_{\textrm{BZ}}}
\sum_{\sigma \xi}
	\int kdk\Theta(\mu-E^1_{\mathbf{k}\sigma\xi}).
\end{eqnarray}
The total doping
$x$
is equal to
\begin{eqnarray}
\label{eq::constant_doping}
x=\sum_{\sigma \xi} x_{\sigma \xi}.
\end{eqnarray}
If
$\mu > \Delta_{\sigma \xi}$,
we obtain the relation between the partial doping and the chemical potential in the form
\begin{equation}
\label{doping_part}
x_{\sigma\xi}=\frac{3\sqrt{3}}{8\pi}\sqrt{m^2-\delta^2_{\sigma\xi}}.
\end{equation}
Otherwise,
$x_{\sigma\xi}=0$.
As a result, we derive in the case of non-zero
$x_{\sigma\xi}$
\begin{eqnarray}
\label{mu_part}
m&=&\delta_0-\frac{8\pi}{3\sqrt{3}}x_{\sigma\xi}
=
\delta_0 \left( 1 - \frac{2 x_{\sigma \xi}}{x_0} \right),
\\
\label{gap_part}
\delta_{\sigma\xi}&=&\delta_0\sqrt{1-\frac{4x_{\sigma\xi}}{x_0}},
\end{eqnarray}
where
\begin{eqnarray}
x_0=\frac{t_0\Delta_0}{\sqrt{3}\pi t^2}.
\end{eqnarray}
Equation~(\ref{gap_part}) indicates that for
$x_{\sigma \xi} = x_0/4$
the order parameter in the sector vanishes. That is, for
$x_{\sigma \xi} > {x_0}/{4}$
one has
\begin{eqnarray}
\label{eq::MU_paramagnet}
\Delta_{\sigma \xi} (x_{\sigma \xi}) \equiv 0,
\ \
m=\frac{8\pi}{3\sqrt{3}}x_{\sigma\xi}
=
\frac{2 \delta_0}{x_0} x_{\sigma \xi}.
\end{eqnarray}
Note that the chemical potential, as given by
Eqs.~(\ref{mu_part})
and~(\ref{eq::MU_paramagnet}),
demonstrates non-monotonic behavior as a function of
$x_{\sigma \xi}$.
Of particular importance is the fact that, for low doping,
$\mu = \mu (x_{\sigma \xi})$
is a decreasing function. This means that the compressibility of the
homogeneous phase is negative and points to a possibility of the phase
separation of the electronic liquid. We will assume below that the
long-range Coulomb interaction is sufficiently strong to arrest the phase
separation, restoring the stability of homogeneous states.

\section{Quarter metal state of doped AB-BLG}
\label{sec::quarter_metal}

\begin{figure*}
\includegraphics[width=0.22\textwidth]{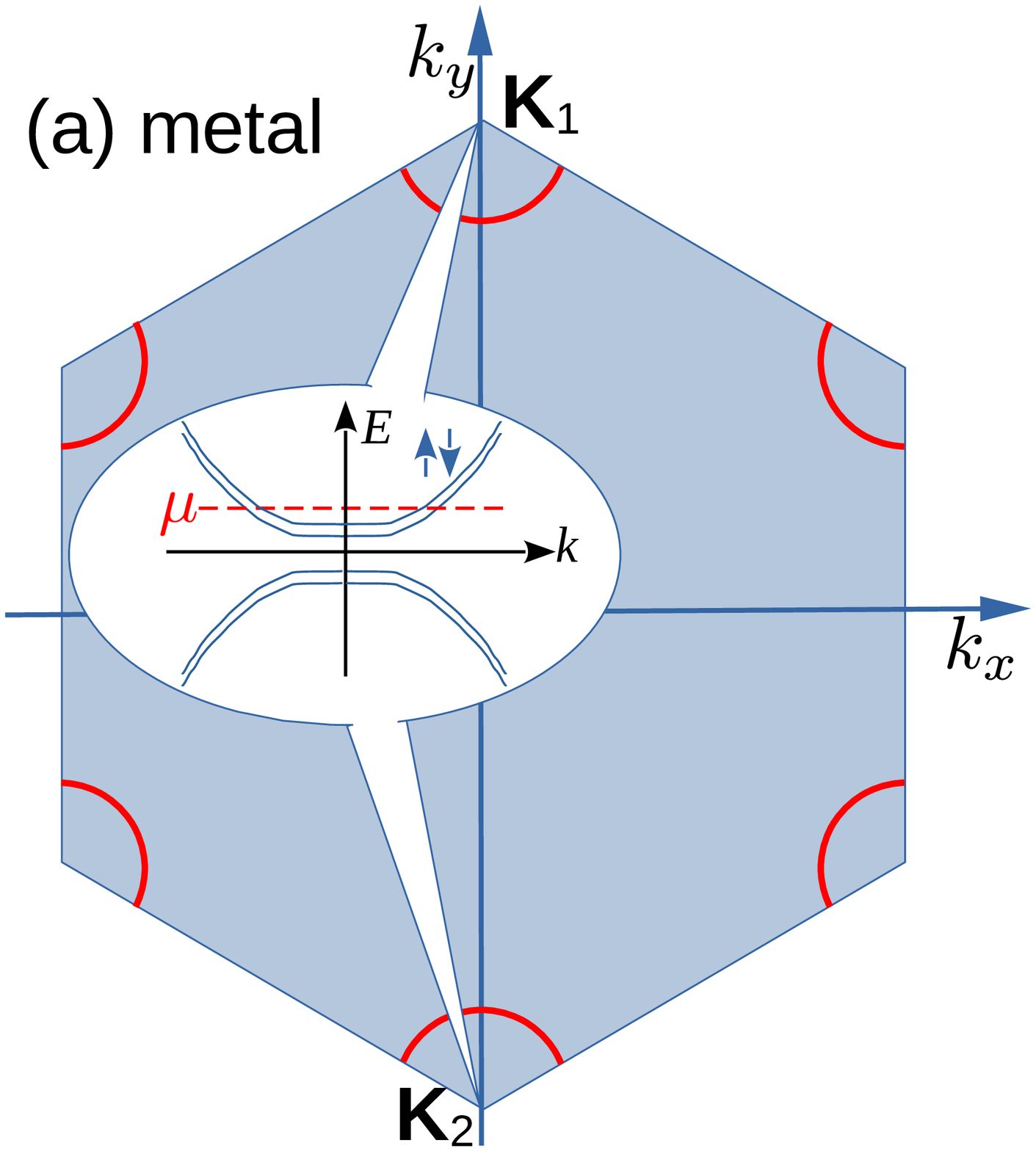}
\hspace{5mm}
\includegraphics[width=0.22\textwidth]{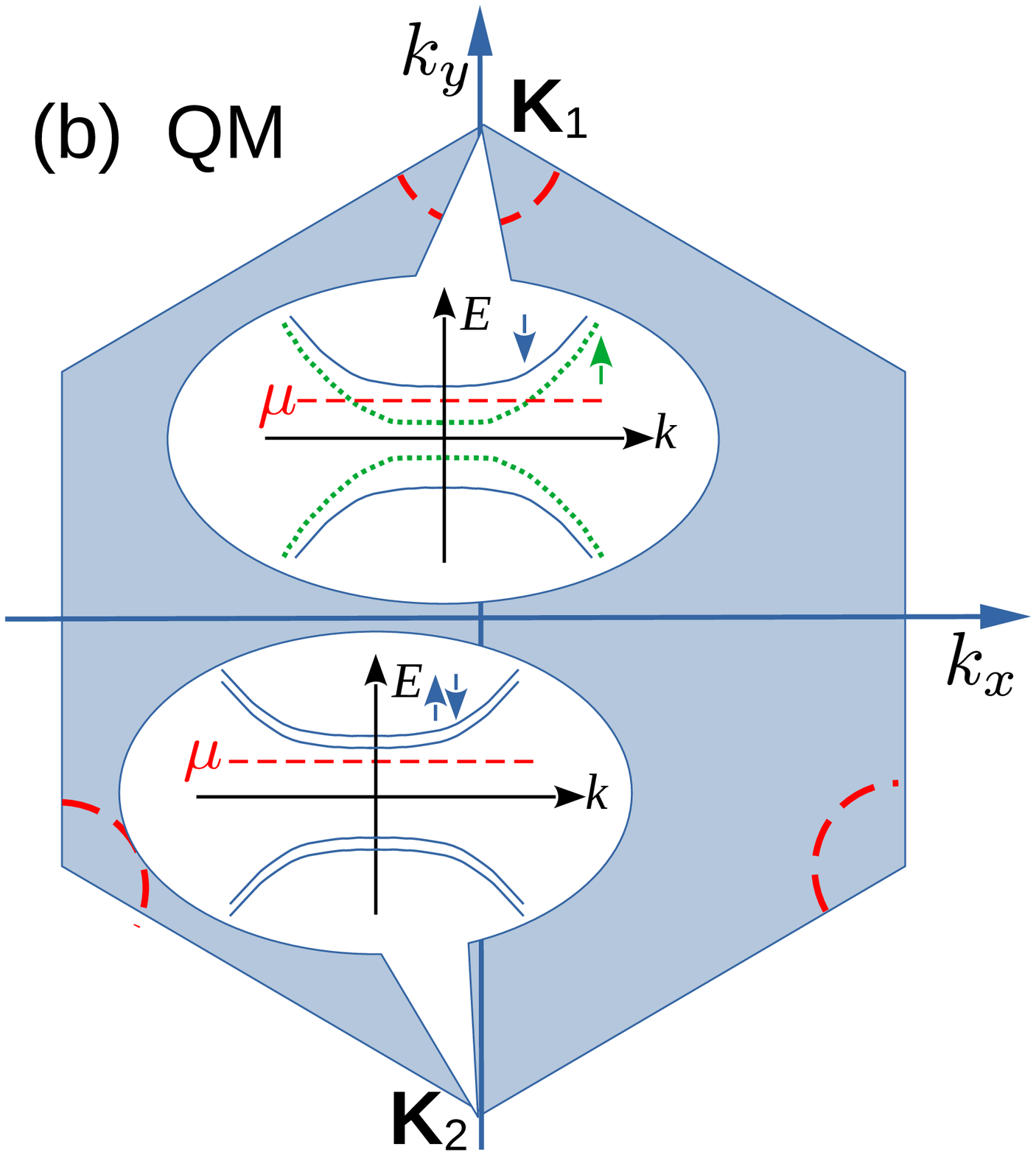}
\hspace{5mm}
\includegraphics[width=0.22\textwidth]{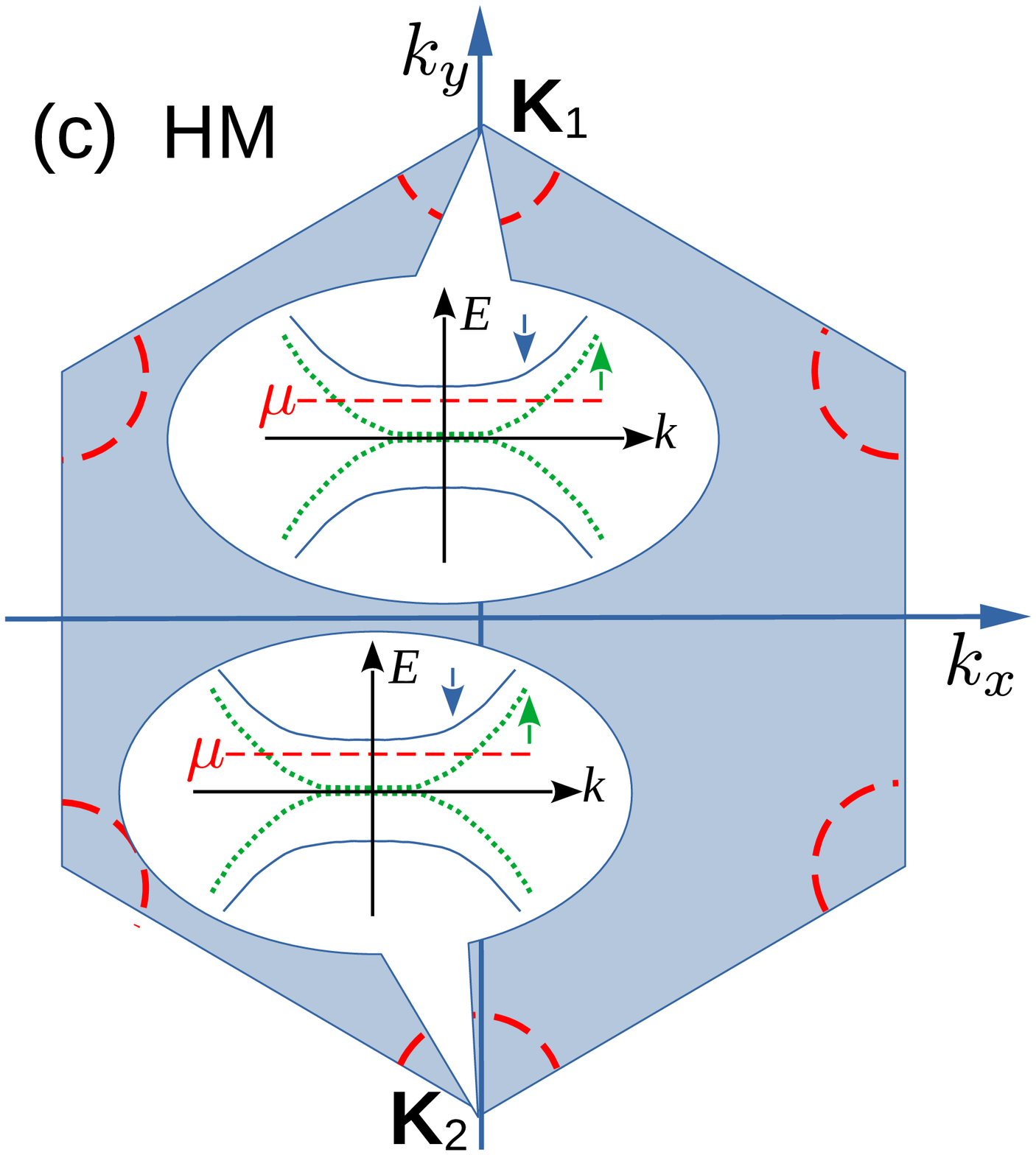}
\hspace{5mm}
\includegraphics[width=0.22\textwidth]{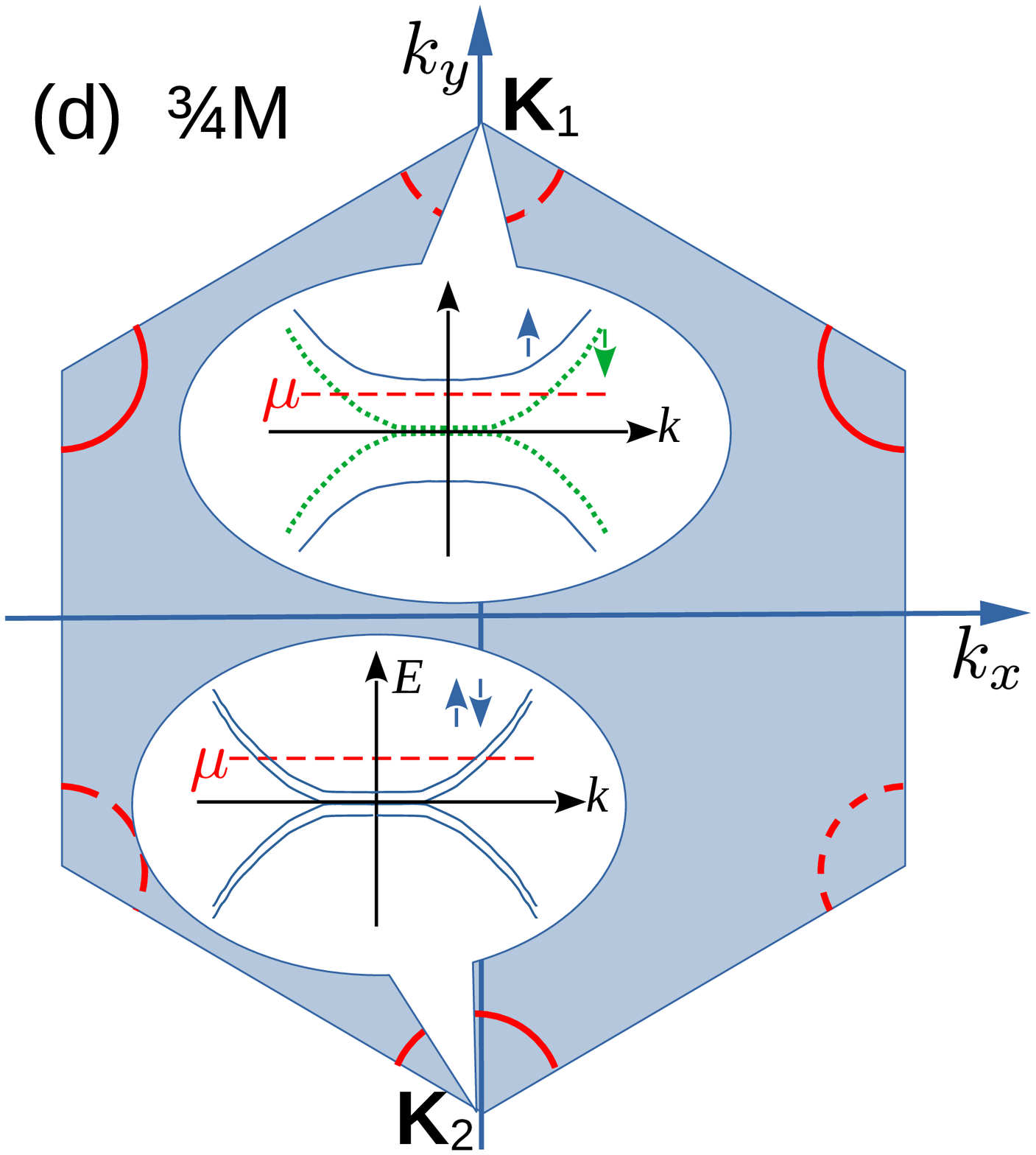}
\caption{
Fermi surface structure of different metallic states. Filled
(blue) hexagon is the Brillouin zone of AB-BLG. Dirac points
${\bf K}_{1,2}$
are marked. Solid and dashed (red) arks near the Dirac points are the Fermi
surface segments. The segments with double degeneracy over the spin-flavor
index are shown as solid curves. Non-degenerate Fermi surface sheets are
represented by dashed arcs.  Diagrams inside callouts depict schematically
the quasiparticle dispersion near a specific Dirac point. Horizontal (red)
dashed line represents chemical potential level. Degenerate bands are shown
by solid (blue) double curves. When this degeneracy is lifted, as in
panels~(b), (c), and~(d), the bands touching or moving closer to one
another, are plotted by dotted (green) curve. Vertical arrows represent the
spin-flavor index
$(-1)^{l+1} \sigma$.
Ordinary metallic state in panel~(a) has a Fermi surface sheet in both
valleys. However, within framework of our model, its energy is higher than
the energy of FraM states (at fixed doping). Panel~(b) depicts the
quarter-metal phase, which is stable at not-too-large doping. For this
state, the available Fermi surface is located in one valley only, and is
non-degenerate. Note that QM is nematic (violates rotation symmetry). A
specific example of a half-metal state is shown in panel~(c). Here the
Fermi surface is present in both valleys, but it is non-degenerate.
Panel~(d) corresponds to 3/4-metal. The Fermi surface is in both valleys,
however, in one valley the Fermi surface sheet is degenerate, in the other
it is not. Because of this, this phase is nematic.
\label{fig::Fermi_surface}
}
\end{figure*}

Disregarding the
possibility of the phase separation, we use
Eqs.~(\ref{mu_part})
and~\eqref{gap_part}
to characterize the thermodynamics of the system. To describe the doped
state of the electronic liquid for a specific $x$, one must determine
partial dopings in all four sectors. To achieve this goal, we calculate the
free energy
\begin{eqnarray}
\label{eq::total_free_en}
F(x) = F(0) + \sum_{\sigma \xi} \delta F(x_{\sigma \xi}).
\end{eqnarray}
In this formula
$F(0)$
is the free energy of the undoped system, and
$\delta F (x_{\sigma\xi})$
shows how much the sector
($\sigma$,~$\xi$)
contributes, for given partial doping
$x_{\sigma\xi}$,
to the total free energy
$F(x)$.
The contribution
$\delta F (x_{\sigma \xi})$
can be found with the help of the relation
\begin{eqnarray}
\delta F (x_{\sigma \xi}) = \int_0^{x_{\sigma \xi}} \mu(x')dx',
\end{eqnarray}
and
Eq.~(\ref{mu_part})
and~(\ref{eq::MU_paramagnet})
that connect the chemical potential and partial doping. Thus we derive
\begin{eqnarray}
\label{free_en_x}
\delta F(x_{\sigma \xi})
=
\begin{cases}
	\Delta_0 \left( x_{\sigma \xi} - \frac{x_{\sigma \xi}^2}{x_0} \right),
	& \text{if } 0 \leq x_{\sigma \xi} \leq \frac{x_0}{4},
	\\
	\Delta_0
	\left( \frac{x_0}{8} + \frac{x_{\sigma \xi}^2}{x_0} \right),
	& \text{if } x_{\sigma \xi} > \frac{x_0}{4}.\\
\end{cases}
\end{eqnarray}
The free
energy~(\ref{eq::total_free_en})
must be minimized over
$x_{\sigma \xi}$
under the
constraint~(\ref{eq::constant_doping}).

For a generic value of $x$, the particulars of such a minimization
procedure might be somewhat cumbersome. Yet for small doping
$x < x_0/4$,
calculations simplify significantly due to all partial dopings being
limited by
$x_0/4$
from above. In this regime one can demonstrate that $F$ is smallest when
all charges are placed into a single sector
\begin{eqnarray}
\label{eq::QM_doping}
x_{\sigma \xi} = x,
\quad
x_{\sigma' \xi'} = 0,
\text{ for }
\sigma' \ne \sigma
\text{ or }
\xi' \ne \xi.
\end{eqnarray}
For the
distribution~(\ref{eq::QM_doping}),
the doping-dependent part of the free energy equals to
\begin{eqnarray}
F_{\rm QM}  = \Delta_0 \left(x - \frac{x^2}{x_0} \right).
\end{eqnarray}
It is smaller, for example, than the free energy
\begin{eqnarray}
F_{\rm eq}  = \Delta_0 \left(x - \frac{x^2}{4x_0} \right)
\end{eqnarray}
calculated for an equal distribution of doping between all four sectors
($x_{\sigma \xi} = x/4$
for all $\sigma$ and $\xi$).

The state described by
Eq.~(\ref{eq::QM_doping})
is metallic, with (almost) circular Fermi surface whose radius
$k_{\rm F} = k_{\rm F} (x)$
is set by the equation
\begin{eqnarray}
a_0^2 k_{\rm F}^2 = \frac{8\pi x}{3\sqrt{3}}.
\end{eqnarray}
This Fermi surface, however, is quite unique: all single-electronic states
reaching the Fermi energy are perfectly polarized in terms of
$S^{\rm f}$
and
$S^{\rm v}$.
In other words, they have an identical value of
$(-1)^{l+1} \sigma$,
and the Fermi surface is located within a single valley
${\bf K}_\xi$.
Since among four possible Fermi surface sheets of the non-interacting
theory, only one sheet emerges in the system, it is natural to designate
such a conducting state as a QM. To appreciate the difference between a
metal with equal distribution of charges between the sectors on one side
and a QM on the other side, one can compare pages~(a) and~(b) of
Fig.~\ref{fig::Fermi_surface}.

\section{Cascade of phase transition between different symmetry-broken
phases}

The QM state described above remains stable only for sufficiently low $x$:
one sector cannot accommodate too much doping. Indeed, when
$x = x_0/2$,
Eq.~(\ref{eq::MU_paramagnet})
implies that
$\mu = \Delta_0$.
Doping a single sector beyond this point is impossible: adding more charge
to this sector increases the chemical potential beyond
$\Delta_0$,
unavoidably placing charges into the remaining sectors as well. As a
result, a cascade of doping-driven phase transitions emerges. The
transitions connect different metallic states, each state being
characterized by a number of doped sectors:
$1,\,2,\,3$,
or $4$ [paramagnetic (PM) state] sectors.

\begin{figure*}
\includegraphics[width=0.6\textwidth]{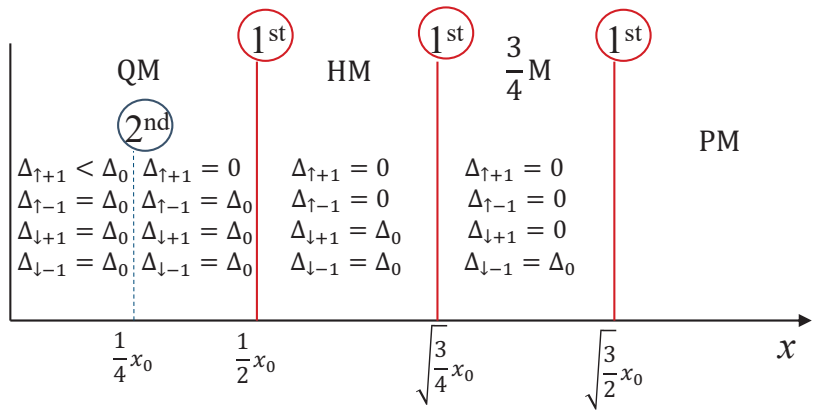}
\caption{Cascade of the doping-driven phase transitions between different
FraM states with different valley and/or spin-flavor (isospin)
polarizations. Only the region of electron doping is shown. For hole doping
the picture is identical up to a replacement
$x\to-x$.
Vertical solid (dashed) lines represent first (second) order transitions.
\label{fig::diagram}
}
\end{figure*}

Let us briefly describe this cascade of transitions (see
Figs.~\ref{fig::Fermi_surface}
and~\ref{fig::diagram}).
At zero doping the system is gapped with the gap equal to
$\Delta_0$
in all sectors. For small $x$, the system absorbs all extra charge
carriers into a single sector [say, sector
($\sigma=\uparrow$,
$\xi=+1$)].
This is a QM state
[Fig.~\ref{fig::Fermi_surface}(b)].
The order parameter in this sector gradually decreases with doping. At the
same time, the chemical potential decreases with doping indicating the
possibility of the phase separation. However, we assume that the long-range
Coulomb repulsion totally arrests the phase separation and the electronic
state remains homogeneous. At
$x=x_0/4$,
the order parameter in doped sector vanishes, and a second order phase
transition inside the QM state takes place. This transition is
characterized by the complex order parameter and a presence of the
developed Fermi surface.

Beyond
$x=x_0/4$,
order parameter
$\Delta_{\uparrow +1}$
is zero. Yet, the QM state remains stable for
$x<x_0/2$.
At higher doping, the extra charge comes to some other sector [for
definiteness, we assign this to be
($\sigma=\uparrow$,
$\xi=-1$);
other configurations are equiprobable]. However, one can show that the
state when the order parameter of this sector is greater than $0$ but less
than
$\Delta_0$
is metastable one. The stable state corresponds to
$\Delta_{\uparrow-1}=0$.
As a result, there appears a first order phase transition between QM state
with
$\Delta_{\uparrow +1}=0$
(other sectors are gapped) and HM state with
$\Delta_{\uparrow +1}=\Delta_{\uparrow-1}=0$
(other sectors are gapped)
[Fig.~\ref{fig::Fermi_surface}(c)].
It happens at
$x=x_0/2$.
This critical doping is found by comparison of the free energies of
corresponding states.

As $x$ increases further, one reaches the point where the HM energy becomes
equal to that of a $3/4$ metal ($\frac34$M) state [Fig.~\ref{fig::Fermi_surface}(d)]. In such a state, three
sectors
[say,
($\sigma=\uparrow$, $\xi=+1$),
($\sigma=\uparrow$, $\xi=-1$),
and
($\sigma=\downarrow$, $\xi=+1$)]
are doped, and the fourth sector,
($\sigma=\downarrow$,
$\xi=-1$),
is gapped, with the extra charge carriers being equally distributed among
the three doped sectors. Again, one can show that the state corresponding
to
$0<\Delta_{\downarrow+1}<\Delta_0$
is metastable one. In the stable
$\frac34$M
state the order parameters in all three doped sectors vanish. As a result,
there appears a first order phase transition between HM and
$\frac34$M
states. Comparing the free energies of these two states, one
finds the point of the transition. It appears at
$x=\sqrt{3/4}x_0$.

If doping is continued even further, the
$\frac34$M
state is replaced by the PM state
[Fig.~\ref{fig::Fermi_surface}(a)].
This is yet another first-order transition, and the last one in the
transition cascade. It occurs at
$x=\sqrt{3/2}x_0$.
The value of this doping is found by comparison of the free energies of
$\frac34$M
and PM states. The phase diagram of the system is shown in
Fig.~\ref{fig::diagram}.
In this figure only the electron doping is shown. Due to electron-hole
symmetry of our model, the phase diagram at hole doping is equivalent to
that shown in
Fig.~\ref{fig::diagram}
up to the replacement
$x\to-x$.

\section{Discussion}

We would like to stress here several important
points. One must remember that the HM state realized in our model upon
sufficiently strong doping is not the conventional
HM~\cite{first_half_met1983,half_met_review2008}
whose Fermi surface demonstrates perfect spin polarization. Instead, we now
have a spin-flavor
HM~\cite{our_hmet_prl2017, our_hmet_prb2018,
jetp_lett_minireview_half_met2020, neutrons_HM2020},
with perfect spin-flavor polarization of the Fermi surface. This means that
the electron (hole) single-particle states reaching the Fermi energy have
their spin projection being equal to
$\sigma$
(to
$\bar \sigma$).
(The related feature of the QM state was already mentioned above.)
In a model with electron-hole symmetry a
spin-flavor-polarized FraM state does not accumulate net spin polarization.
However, a finite spin polarization may accompany a finite spin-flavor
polarization~\cite{our_hmet_prl2017}
when such a symmetry is absent. The spin polarization was indeed observed
in
Ref.~\onlinecite{ab_frac2022exper}.

We argued above that the relative stability of various metallic states is
affected by doping, triggering the transitions between them. Doping is not,
however, the only factor that influence the competition between the FraM
phases. Particular model's ingredients favoring HM states are the umklapp
and backscattering interaction terms. Specifically, the umklapp couples two
sectors with unequal
$(-1)^l \sigma$,
the backscattering, on the other hand, connect the sectors with
non-identical values of the $\xi$ index. Thus, in the presence of either
strong
$H^{\rm u}_{\rm int}$
or strong
$H^{\rm b}_{\rm int}$
only two (not four) decoupled sectors of the mean-field Hamiltonian can be
defined, promoting the HM phase over other FraM's. Therefore, in more
realistic models, the critical doping values are no longer proportional to
$x_0$,
with universal proportionality coefficients. Instead, they become functions
of the backscattering and umklapp coupling constants. Finally, one must
remember that our single-electron Hamiltonian is based on the simplest
effective model of AB-BLG. It unavoidably ignores some details of the
AB-BLG band structure, such as the trigonal warping caused by a
longer-range hopping
terms~\cite{mccann_falko2006,jung_mcdonald2014tb}.
Specifically, the trigonal warping acts to replace the parabolic dispersion
of the
Hamiltonian~(\ref{Ham_1})
with four Dirac cones, depleting the density of states at the Dirac
points. The latter, in turn, reduces the transition temperature, making the
transition itself even more dependent on the strength of the interaction.
Fortunately, there is ample experimental evidence suggesting
that electron-electron interaction in AB-BLG is sufficiently strong to
cause low-temperature ordering. Thus, as the simple approximation, these
band effects can be ignored, and
Hamiltonian~(\ref{Ham_1})
can be used. Yet, for more detailed modeling of the transition cascade
a more accurate band description is necessary.

The qualitative agreement between the remarkable recent experiments
reported in
Ref.~\onlinecite{ab_frac2022exper}
and our formalism is very encouraging. The proposed theory can account for
such experimentally observed features as the cascade of phase transitions,
magnetization, and valley polarizations. Yet one must keep in mind that the
experiments were performed at finite electric field applied transverse to
a sample. In our formalism, this field is assumed to be zero. Further
research is needed to understand the role of this field.

To conclude, we proposed a mechanism responsible for the formation of the
FraM states in doped AB-BLG. We argue that, as doping increases, this
system demonstrates a cascade of phase transitions between various metallic
phases that differ in terms of spin-flavor and valley polarizations of
their Fermi surfaces. Our theoretical findings compare
favorably to very recent
experiments~\cite{ab_frac2022exper}
on AB-BLG.


\end{document}